\documentclass[aps,pra,twocolumn,reprint,groupedaddress, nofootinbib]{revtex4-1} 
\usepackage{graphicx}
\usepackage{amsmath}
\usepackage{epstopdf}
\begin{document}
\title{Generation of frequency sidebands on single photons with indistinguishability from quantum dots}
\author{Uttam Paudel}\altaffiliation{Now at The Aerospace Corporation, ElSegundo, California, USA}
\email[]{upaudel@umich.edu}
\affiliation{Department of Physics at the University of Michigan, Ann Arbor, Michigan, USA}
\author{Alexander P. Burgers}\altaffiliation{Now at Norman Bridge Laboratory of Physics, California Institute of Technology, Pasadena, California, USA}
\affiliation{Department of Physics at the University of Michigan, Ann Arbor, Michigan, USA}
\author{Duncan G. Steel }
\affiliation{Department of Physics at the University of Michigan, Ann Arbor, Michigan, USA}
\author{Michael K. Yakes}
\affiliation{Naval Research Laboratory, Washington DC, USA }
\author{Allan S. Bracker}
\affiliation{Naval Research Laboratory, Washington DC, USA }
\author{Daniel Gammon}
\affiliation{Naval Research Laboratory, Washington DC, USA }
\date{\today}

\begin{abstract}
Generation and manipulation of the quantum state of a single photon is at the heart of many quantum information protocols. There has been growing interest in using phase modulators as quantum optics devices that preserve coherence. In this Letter, we have used an electro-optic phase modulator to shape the state vector of single photons emitted by a quantum dot to generate new frequency components (modes) and explicitly demonstrate that the phase modulation process agrees with the theoretical prediction at a single photon level. Through two-photon interference measurements we show that for an output consisting of three modes (the original mode and two sidebands), the indistinguishability of the mode engineered photon, measured through the second-order intensity correlation ($g^2(0)$) is preserved. This work demonstrates a robust means to generate a photonic qubit or more complex state (e.g., a qutrit) for quantum communication applications by encoding information in the sidebands without the loss of coherence. 
\end{abstract}


\maketitle

Quantum communication and computing protocols often require a flexible and customizable single photon source that can form a link between distant nodes \cite{Kimble, bennett, divinc}. Swapping entanglement between these nodes can be achieved utilizing the two-photon interference (HOM) measurements \cite{monroe, grangier, ou}, where the optimal interference to assure indistinguishability requires that the spatial, temporal, polarization, and spectral modes of the input photon wavefunctions must be identical \cite{ou, Santori_2002}. Thus manipulation of the photonic degrees of freedom while maintaining coherence is very important for many quantum information applications.

Single photons also function for cryptographic key distributions to enable transfer of information between two remote parties \cite{bennett, Bennett_2014}, where quantum information is encoded in the various degrees of freedom of a single photon. Polarization qubits are typical, but are prone to decoherence when transmitted through a fiber \cite{Tomita1986, Ross_1984, Breguet}. Frequency qubits \cite{Duan_2006}, on the other hand, are known to be robust against any mechanically or environmentally induced fluctuation in a fiber \cite{qkdphase, Lukens, ramseyphoton}. Frequency qubits can be generated through phase modulation of a single photon \cite{Lukens, EOMentanglement}, where the information is encoded in the relative amplitude between the sidebands.  Recently, Lukens and Lougovski proposed a universal linear-optical quantum computing (LOQC) platform using frequency components generated from an electro-optic modulators \cite{Lukens}. Similarly, there has been proof-of-concept demonstrations of the BB84 protocol using phase modulated weak coherent sources \cite{qkdphase, mora}. 

A quantum analysis of phase modulation was first discussed by Louisell, Yariv, and Siegman \cite{louisell}. Frequency conversion is described as a two-mode coupling process with sinusoidal coupling between the unmodulated and the new frequency component. The coupling between the two modes is generated through a periodic perturbation of the refractive index of the medium \cite{louisell}. Building on that work, Miroshnichenko et al. explicitly derive a multimode Hamiltonian for a modulator and give a fully quantized description of the phase modulation with single photons, where the field amplitudes at different frequency components are weighted by the Bessel coefficients \cite{Miroshnichenko}. The work by Kolchin et al. \cite{Kolchin} lays an experimental foundation for the pulse shaping at a single photon level and has motivated several works on the topic \cite{harris_2009, harris_2010, Fan, Wright, karpinski16}. The timeliness for a demonstration is highlighted by the recent publication of papers that use a phase modulator as a quantum optics device that preserves quantum coherence \cite{eom_jianwei, eom_imamoglu, kues_2017, lo_2017, lu_2018}.

In this paper,   we build on the work above and use single photons from an isolated InGaAs quantum dot to show agreement with the quantum analysis \cite{Miroshnichenko} when  an electro-optic phase modulator is used to produced frequency side bands centered around the emission frequency of the dot.  A Hong Ou Mandel (HOM) interferometer is used to demonstrate preservation of quantum coherence in the resulting quantum coherent superposition state.  The measurements demonstrate a robust means of manipulating a photonic qubit.

A single exciton in an InAs/GaAs quantum dot (QD) nanostructure behaves very similarly to a two-level system \cite{Stievater_2001}, where the single photons emitted by the nanostructures can form excellent photonic qubits \cite{imamoglu}. By resonantly exciting a single QD with a continuous-wave (CW) laser, one can generate a stream of single photons with up to a GHz emission rate \cite{Stock_2010}. When the exciton is resonantly driven with a weak excitation laser (the Rabi frequency is much less than the transition linewidth), it is shown that the coherence time of the scattered single photons are the same as the excitation laser with the mutual coherence between the laser and the scattered photon $\sim3$ seconds \cite{heitler, mete_phase}. This makes QDs a unique source for generating an ultra-bright stream of photons with high single photon purity ($g^{(2)}\sim 0$) and a long coherence time for quantum communication applications that otherwise are not possible with the down-conversion source or from an atomic system \cite{Stock_2010, shields_2006, mete_phase, Somaschi, jianwei_2016}.

A resonantly driven two-level system with a quasi-monochromatic coherent source in the absence of decay beyond spontaneous emission scatters the incident field elastically (Rayleigh scattering) \cite{mollow_1969}. The scattered field inherits the properties (including the spectral bandwidth) of the incident field while exhibiting sub-Poissonain statistics \cite{mete_phase, Kimble1977}, which we measure in a Hanbury Brown and Twiss (HBT) interferometer with two fast single photon detectors \cite{Kimble1977}. 
For a two-level system in the rotating wave approximation, the second-order intensity correlation function for arbitrary field strength is given by \cite{scully_QO},
\begin{widetext}
\begin{equation}
g^{(2)} (\tau) = \lim_{t \to \infty}\frac{\langle I(t)I(t+\tau)\rangle}{\langle I(t)\rangle^2 } = 1- [cos(\mu |\tau|) + \frac{\gamma+\gamma_2}{2\mu} sin(\mu|\tau|)]e^{-\frac{1}{2}(\gamma+\gamma_2)|\tau|}
\label{eq:g2theory02}
\end{equation}
\end{widetext}
with $\mu = \sqrt{\left(\Omega^2_0-\frac{(\gamma-\gamma_2)^2}{4}\right)}$ where $I(t)$ is the intensity of the field detected by a detector at time t, $\tau$ is the relative time difference between the two detectors,  $\Omega_0$ is the Rabi frequency and $\gamma_2$ and $\gamma$ are the spontaneous decay rate and dephasing rate of the QD, where the lifetime is given by $(2\pi\gamma_2)^{-1}$. Physically, $g^{(2)} (\tau)$ is the probability of detecting a photon at $t=\tau$ if a photon was detected at $t=0$.   For an ideal single photon source,  $g^{(2)}(0)=0$.

\begin{figure}[h]
\begin{center}
\includegraphics [width=0.42\textwidth]{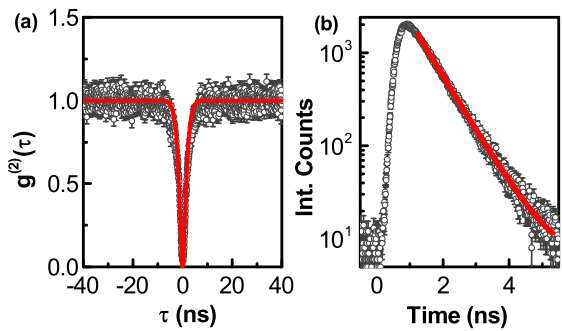} 
\caption{(a) Normalized second-order intensity correlation of a single QD. At $\tau = 0$ the raw coincidence count drops to $g^{(2)} (0)= 0.039 \pm 0.01$, confirming the single photon nature of the emitted stream of photons. The red curve is the theoretical fit to the data obtained using Eq. (1). (b) Semi-log plot of resonantly excited time tagged fluorescence emission of a single QD. The red curve is a single exponential with an offset fitted to the data, which gives the emission lifetime to be $745\pm 5$ ps. All error bars plotted in the paper are standard error of the mean. 
}
\end{center}
\end{figure}

In this Letter, we use a resonantly excited, self-assembled InAs/GaAs QD with a narrow linewidth CW laser in order to generate a stream of bright single photons which are detected in a cross-polarization setup. The QDs are embedded in an asymmetric distributed Bragg reflector (DBR) cavity with a small Q factor of $\sim90$. We can detect a raw single photon count rate of 1.25 million per second with the QD in study.
Figure 1(a) is the measured raw data and a theoretical fit using Eq. (1) for the second-order intensity correlation with a resonantly excited single QD. The fit gives the radiative lifetime to be $695\pm50$ ps. At $\tau = 0$, $g^{(2)} (0) = 0.039 \pm 0.01$, and is limited by the detector timing resolution. This indicates the QD in the study is an excellent single photon source. Such a source with vanishingly small multi-photon emission rate would eliminate the well-known photon-number splitting attack an eavesdropper could otherwise use on a coherent state \cite{PNS,PNS2}.

Figure 1(b) shows the radiative lifetime data from a resonantly excited single QD. The QD in study is excited with 50 ps pulses and the time resolved emission histogram is built by syncing the detected photons with the excitation pulse \cite{Schaibley_2013}. The red curve is an exponential fit to the data, which gives the emission lifetime to be $745 \pm 5$ ps, consistent with the lifetime extracted from the $g^{(2)}$ measurement.

\begin{figure*}
\includegraphics[width=0.78\textwidth]{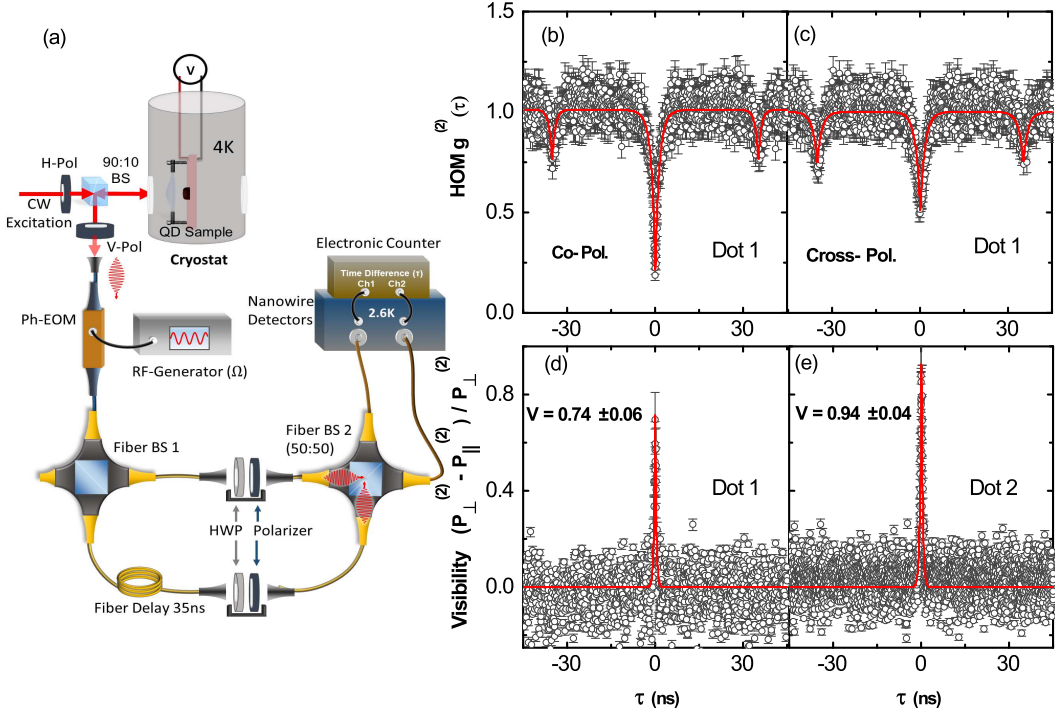} 
\caption{(a) Experimental setup for the two-photon interference measurements. The quantum dot is resonantly excited by a CW-laser and the collected photons are sent through a fiber phase modulator (turned off for this measurement) and an unbalanced fiber Mach-Zehnder interferometer. The same setup with the first BS removed is used to measure the statistics of the emitted photons as seen in Fig. 1(b). (b) Two-photon interference measurements with dot 1 for linearly co-polarized and (c) linearly cross-polarized photons with the phase modulator off. For co-polarized photons, the $g^{(2)}_{\text{HOM}}(\tau=0)$  goes to $0.19\pm.03$. In the case of the cross-polarized photons, the value is 0.5, which is the classical correlation value. The two side dips at tau $\tau = \pm 35$ ns correspond to the relative path length difference of the Mach-Zehnder interferometer where the coincidence count is reduced due to the classical counting probability for single photon source, and not due to the interference. Therefore the magnitude of these dips are equal for the co-polarized and cross-polarized cases. Figure (d) and (e) are interference visibilities measured for dot 1 and dot 2. The red curve on plot b, c, and d-e are obtained from the theoretical fit given by Eq. (2), (3) and (4) respectively.}
\end{figure*}

To show the indistinguishability of the single photons, we perform continuous-wave Hong-Ou-Mandel type two-photon interference measurements \cite{patel, proux} in an unbalanced fiber Mach-Zehnder interferometer by exciting the QD near saturation. One arm of the interferometer is delayed 35 ns relative to the other arm such that two photons meet at the second beam splitter simultaneously. The long delay ensures there is no interference due to the first-order coherence (i.e., $g^{(1)}$). The experimental setup for the HOM measurement is given in Fig. 2(a).

When the incident photons are identical in all degrees of freedom, the input photons exit through the same port of the beam splitter (BS2 in Fig. 2(a)). This results in a drop in coincidence counts between the two detectors. The HOM interference measurement can be calculated with the second-order intensity correlation function. The normalized coincidence probability for the HOM measurements are given by \cite{patel}
\begin{equation}
P^{(2)}_{\parallel} (\tau) = \frac{1}{2}g^{(2)}(\tau) + \frac{1}{4}[g^{(2)}(\tau-\Delta \tau)+g^{(2)}(\tau+\Delta \tau)](1-v_c e^{-2\frac{|\tau|}{\tau_c}})
\label{eq:g2homcw}
\end{equation}
\begin{equation}
P^{(2)}_{\bot} (\tau) = \frac{1}{2}g^{(2)}(\tau) + \frac{1}{4}[g^{(2)}(\tau-\Delta \tau)+g^{(2)}(\tau+\Delta \tau)]
\label{eq:g2homcw2}
\end{equation}
where $P^{(2)}_{\parallel}$ and $P^{(2)}_{\bot} $ correspond to the cases where the input photons in BS2 are linearly co-polarized and linearly cross-polarized respectively and $g^{(2)}$ is the intensity correlation function given in Eq.(1). The two beam-splitters (BS1 and BS2) used in the setup are $50:50$ (R:T) and polarization insensitive, $\Delta \tau$ is the relative time difference between the two arms of the interferometer, $\tau_c$ is the coherence time of the photons, and $v_c$ is the overlap of all the modes between the two incident photons that takes the value from 0 (no overlap) to 1 (perfect overlap). The overall indistinguishability of the photons is quantified by a visibility function defined as \cite{patel}, 
\begin{equation}
V_{\text{HOM}}(\tau) = \frac{P^{(2)}_{\bot} - P^{(2)}_{\parallel}}{P^{(2)}_{\bot}}
\label{eq:cwvis}
\end{equation}
For completely indistinguishable photons, the visibility goes to 1 at $\tau = 0$, whereas it goes to 0 for distinguishable photons.

Figure 2(b) and 2(c) are the normalized raw HOM data for dot 1 performed with the phase modulator off for linearly co-polarized and cross-polarized photons respectively. The red lines are the theoretical fit to the data obtained from Eq.(2-3).  At $\tau = 0$, for the co-polarized case, the normalized coincidence counts drops to $0.19\pm.03$ (Fig. 2(b)), much below the classical correlation limit of 0.5, whereas for the orthogonal case, the coincidence rises to $0.5\pm.03$ (Fig. 2(c)). We obtain a visibility of $0.74\pm.06$  with the dot (dot 1) used in the study. A different dot (dot 2)\footnote{Experimental parameters (such as the scanning etalon) for this study  were optimized for dot 1; however, we reported data from dot 2 due to its high visibility.} from the same sample has a HOM visibility of up to $0.94\pm.04$  (Fig. 2(e)). This is the highest raw visibility seen in CW HOM measurements reported to date. The reduction in visibility (Fig. 2(d)) from the ideal case of unity is likely due to the spectral diffusion of the QD emission frequency, and uncontrolled polarization rotation of the fiber before the second beam splitter. 

To demonstrate the ability to alter the frequency state of a stream of single photons emitted from an isolated QD, we resonantly excite the QD with a CW laser as in the previous section and the scattered light is passed through an electro-optic phase modulator driven by a sinusoidal microwave field with frequency $\Omega$. The state-vector of a single photon becomes modified with additional frequency modes that were previously unoccupied. The new frequency components are separated by the harmonics of the microwave field with amplitudes determined by the microwave field modes coupling with the optical field \cite{Miroshnichenko}. For a monochromatic input field ($|\psi_{\text{in}}\rangle = |1_{\omega_0}\rangle$) of the single photon centered at $\omega_0$, at the limit that a large number of microwave modes are coupled with the optical field, the output state can be expanded as an infinite sum of the Bessel coefficients of the first kind \cite{eomclassical,Miroshnichenko},
\begin{equation}
|\psi_{\text{out}}\rangle = \sum_{n=-\infty}^{\infty} J_n(\beta) e^{i(\theta-\pi/2)n} |1_{\omega_0+n\Omega}\rangle
\label{eq:out}
\end{equation} 
where $\beta$ is the modulation index and $\theta$ is the phase of the microwave field. See the Supplementary for details.
\begin{figure}[htb!]
\includegraphics[width=0.45\textwidth]{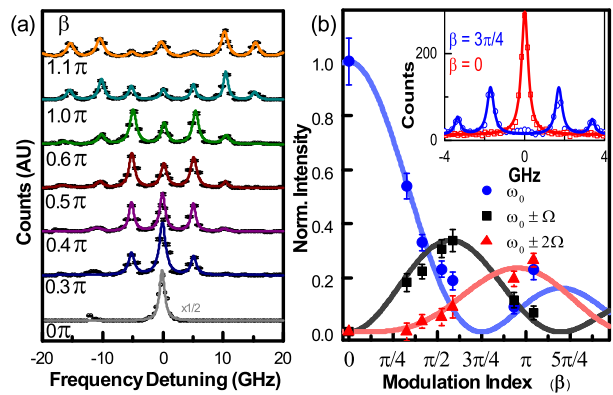} 
\caption{(a) Scanning Fabry-P\'{e}rot spectrum of single photons after sending through the electro-optic phase modulator for various modulation indices ($\beta$). The solid lines are the Lorentzian fit to the data. The phase modulator is driven with a 5 GHz microwave-driver. The relative variance of the intensity of the sidebands is due to the finite discrete stepping resolution of the scanning etalon. (3b) Integrated counts as a function of modulation index for the central peak and the average of the first two sidebands obtained for the fit. The data is normalized by taking the integrated count for the unmodulated case to be one. The solid lines are the square of the zeroth, first and second order Bessel coefficients plotted as a function of the modulation index. (3b-inset) Shows suppression of the carrier component within the extinction contrast with a modulator driven at $3\pi/4$ modulation index. }
\end{figure}

Figure 3(a) is the intensity profile of the phase modulated single photons emitted by a QD driven resonantly at the saturation point. The bottom curve is the unmodulated spectrum of the single photons emitted by the QD. The remaining curves show the modified spectrum of a single photons after the phase modulator, driven by a 5 GHz microwave field at various modulation indices. The total integrated counts remain constant for all modulation indices. The data is fitted with multiple Lorentzian peaks with weighted coefficients. Figure 3(b) shows the integrated counts for the central peak $(\omega_0)$ and the average of the first two sidebands ($\omega_0\pm\Omega$ and $\omega_0 \pm 2\Omega$) obtained from the fit for various modulation indices ($\beta$). The data are normalized by taking the integrated counts for the unmodulated case to be one. The solid lines are the square of the first three Bessel coefficients ($|J_n(\beta)|^2, n = 0,1,2$) plotted as a function of the modulation index. The fit to the data shows excellent agreement with the theoretical prediction for the phase modulation of a single photon state \cite{Miroshnichenko}. As one can see from Fig. 3(b), the single photons are modulated up to the $\pi$ modulation index. As a particularly important example, Fig. 3(b) (inset) shows suppression of the carrier component---for a different quantum dot and 1.7 GHz drive frequency---to within the extinction contrast, when the single photons are modulated at $3\pi/4$ index. The bandwidths of the sidebands are the same as the bandwidth of the unmodulated carrier field, within the margin of error, limited by the finesse of the scanning Fabry-P\'{e}rot etalon. The sidebands are generated at the harmonics of the driving field. The linewidth of the primary and sidebands are given by the excitation laser linewidth when the scattering process is dominated by elastic scattering; however, in these experiments we are operating at powers where the Rabi frequency is close to the natural linewidth.  Here, Mollow shows the line broadens considerably compared to the near delta-function behavior associated with quasi-monochromatic excitation \cite{mollow_1969}.
\begin{figure}[tb]
\includegraphics[width=0.4\textwidth]{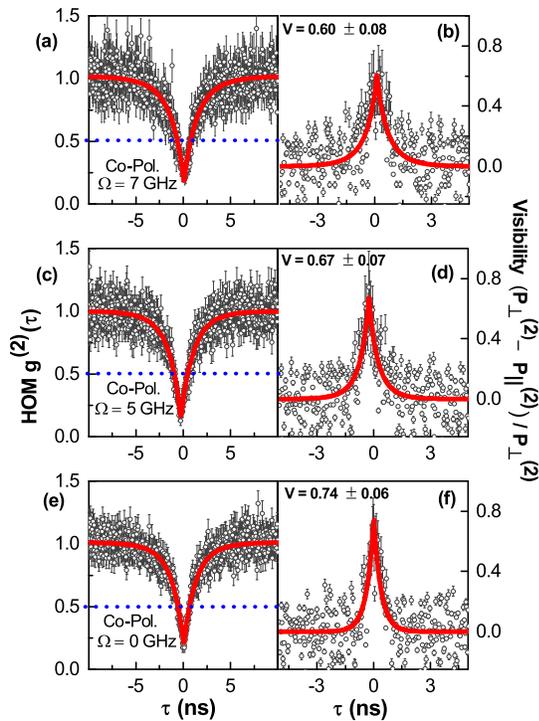} 
\caption{Two-photon interference measurements for linearly co-polarized photons modulated at (a) 7 GHz (c) 5 GHz  and (e) unmodulated case. The interference visibility for each driving frequency is plotted with the same order in the right column. For Fig. (a-d), the modulator is driven at $\beta \sim \pi/3$ index and all frequency components are used in the HOM measurements. The red curves are obtain from the theoretical fit given by Eq. (2-4) for the unmodulated case. The dashed blue lines in the left figures are the classical correlation limit, the normalized coincidence counts below the line indicates the quantum interference between the two photons. The interference visibility, thus the indistinguishability is well preserved for the modulated photons.}
\end{figure}

After the QD photons pass through the phase modulator, it is important to verify that the photon indistinguishability is not degraded substantially by the phase modulation process. If the modulation process destroys the relative phase information between sidebands at the single photon level, or if the sideband generation happens as a statistical process, the effect of them would be manifested as beating in the coincidence counts in a EOM HOM experiment \cite{legero04, rempe_2009}. Due to the finite detector resolution used in the experiment, the oscillations would be washed-out resulting in a substantial increase of the two detectors firing simultaneously ($P_{12}(\tau=0)$) for co-linearly polarized photons. In contrast, if the frequency components are generated in a coherent superposition and the relative phase information between sidebands are preserved, the input photons would be identical from shot to shot, resulting in high interference visibility \cite{imamoglu}.

To see this, we repeat the HOM interference measurements between subsequent photons with the phase modulator turned on and driven by a sinusoidal monochromatic microwave source. The modulation index is $\beta \sim \pi/3$ (blue curve in Fig. 3(a)) where the ratio of carrier mode intensity to the first sidebands is $2.7:1$. Several HOM measurements are performed with driving frequencies ranging from 2 to 7 GHz. The HOM interference for the linearly co-polarized photons with the modulated spectra and the interference visibility are plotted in Fig. 4(a-d). The red curves are the theoretical fit to the data given by Eq. (2-4) for the unmodulated case. Figure 4(e,f) are the detailed plots of 2(b,d), showing HOM interference visibility for the unmodulated photons plotted together for comparison. As seen in Fig. 4(b,d,f), the visibility of the modulated photons remain within the error bars of the unmodulated case and indicates that the modulation process generates additional frequency components as a superposition to the carrier component at a single photon level.

In summary, we have demonstrated that the frequency spectrum of a stream of single photons emitted by a single QD can be modified to generate well-defined frequency sidebands using a phase modulator. We have shown that the sidebands inherit the properties of the unmodulated photon. Using multiple phase modulators in double Mach-Zehnder interferometers \cite{lo_2017, Shimotsu01}, one can actively suppress the unwanted frequency components to construct a photonic frequency qubit $|\psi_{\text{qubit}}\rangle = c_0 |\omega_0\rangle + c_1 |\omega_1\rangle $ \cite{eomclassical, BlochEOM, qkdphase, Lukens, mora}. In addition, there exist proposals to use the carrier component and the first pair of sidebands ($\omega_0, \omega_0\pm\Omega$) as two incompatible bases for BB84 protocols \cite{mora}.  As the modulator is embedded in a single mode fiber, all of the frequency components are in the same spatial mode. This allows transferring such qubit states over a long distance using fiber optic networks without the problems that characterize polarization qubits \cite{Lukens}.
In summary, we have demonstrated that the frequency spectrum of a stream of single photons emitted by a single QD can be modified to generate well-defined frequency sidebands using a phase modulator. We have shown that the sidebands inherit the properties of the unmodulated photon. Using multiple phase modulators in double Mach-Zehnder interferometers \cite{lo_2017,Shimotsu01}, one can actively suppress the unwanted frequency components to construct a photonic frequency qubit $|\psi_{\text{qubit}}\rangle = c_0 |\omega_0\rangle + c_1 |\omega_1\rangle $ \cite{eomclassical, BlochEOM, qkdphase, Lukens, mora}. In addition, there exist proposals to use the carrier component and the first pair of sidebands ($\omega_0, \omega_0\pm\Omega$) as two incompatible bases for BB84 protocols \cite{mora}.  As the modulator is embedded in a single mode fiber, all of the frequency components are in the same spatial mode. This allows transferring such qubit states over a long distance using fiber optic networks without the problems that characterize polarization qubits \cite{Lukens}.
Through HOM measurements, we have demonstrated that the indistinguishability of a stream of individual photons emitted by a QD is fully preserved in the presence of additional frequency sidebands generated via a phase modulator for a range of modulation frequencies.  
These results demonstrate the suitability of this approach for use in the development of frequency qubits from narrow-band single photons and as a basis for a QKD protocol such as BB84.
\begin{acknowledgements}
This work is supported in part by NSF (PHY 1413821), AFOSR (FA9550-09-1-0457), ARO (W911NF-08-1-0487, W911NF-09-1-0406 / Z855204) and DARPA (FA8750-12-2-0333). We would like to thank Aaron Ross for building the scanning Fabry-P\'{e}rot etalon.
\end{acknowledgements}






%

\end{document}



\title{Supplementary Information for ``Frequency sidebands generation of quantum dot single photons with preserved indistinguishability''}


\author{Uttam Paudel}\altaffiliation{Now at The Aerospace Corporation, El Segundo, California, USA}
\email[]{upaudel@umich.edu}
\affiliation{Department of Physics at the University of Michigan, Ann Arbor, Michigan, USA}

\author{Alexander P. Burgers}\altaffiliation{Now at Norman Bridge Laboratory of Physics, California Institute of Technology, Pasadena, California, USA}
\affiliation{Department of Physics at the University of Michigan, Ann Arbor, Michigan, USA}

\author{Duncan G. Steel }
\email[]{dst@umich.edu}
\affiliation{Department of Physics at the University of Michigan, Ann Arbor, Michigan, USA}

\author{Michael K. Yakes}
\affiliation{Naval Research Laboratory, Washington DC, USA }

\author{Allan S. Bracker}
\affiliation{Naval Research Laboratory, Washington DC, USA }

\author{Daniel Gammon}
\affiliation{Naval Research Laboratory, Washington DC, USA }


\date{\today}

\maketitle
Phase modulators use the linear electro-optic effect to modulate the index of the material. When a monochromatic electric field with amplitude $E_0$ and frequency $\omega_0$ is sent through a phase modulator, the input and output fields can be written as
\begin{equation}
 E_{in} = E_0 e^{i\omega_0 t} \hspace{1cm} \text{and}  \hspace{1cm} E_{out} = E_0 e^{i\omega_0 t-i\phi}
\label{eq:phasemodClass1}
\end{equation}
where $\phi=\frac{2\pi}{\lambda} n_x L$  is the phase factor gained by the field of wavelength $\lambda$ traveling through a medium with index $n_x$ and length $L$. When the modulator is driven with a sinusoidal voltage $V = V_m \sin{(\Omega t+ \theta)}$, the index of refraction is modified due to the applied electric field \cite{eomclassical},
\begin{equation}
 \phi=\frac{2\pi}{\lambda}(n_x+\Delta n_x)L
\label{eq:phase}
\end{equation}
For a transverse modulator, where the microwave field is applied transverse to the propagation direction of the optical field, the change in the phase factor becomes \cite{eomclassical}
\begin{equation}
\Delta n_x = -\frac{1}{2} n_x^3 r \left(\frac{L}{d}\right) V_m  \sin{(\Omega t+\theta)},  
\label{eq:index}
\end{equation}
where $r$ is the interaction coefficient and $d$ is the thickness of the crystal across which the voltage is applied. With the modified phase, the output field can be rewritten as 
\begin{equation}
E_{out}=E_0 e^{i\omega_0 t-i\phi_0-i \beta \sin{\Omega t}}
\label{eq:outfield}
\end{equation}
where $\phi_0= \frac{2\pi}{\lambda} n_x L$ is the time-independent phase factor and  $\beta = \frac{\pi}{2} n_x^3 r (\frac{L}{d}) V_m$ is the modulation index, proportional to the applied voltage. The constant phase factor can be absorbed into the electric field amplitude and the exponential can be expanded as a sum of the coefficients of the Bessel functions of the first kind ($J_n$),
\begin{equation}
E_{out}=E_0 \sum^{\infty}_{n=-\infty} J_n(\beta) e^{i n (\theta-\pi/2)} e^{i(\omega_0+n\Omega)t}.
\label{eq:outfield2}
\end{equation}
For notational simplicity, the input field can be represented with ket-notation as $|1_{\omega_0} \rangle$, such that the output state can be written as,
\begin{equation}
E_{out}=\sum^{\infty}_{n=-\infty} J_n(\beta) e^{i n (\theta-\pi/2)} |1_{\omega_0+n\Omega} \rangle.
\label{eq:outfield3}
\end{equation}
A frequency qubit can be formed by suppressing the unwanted frequency components using a single sideband generation technique, where the carrier and a single sideband is generated through the modulation process \cite{Shimotsu01}, 
\begin{equation}
|\psi\rangle=c_0|1_{\omega_0} \rangle + c_1 e^{i\theta}|1_{\omega_0+n\Omega} \rangle.
\label{eq:outfield4}
\end{equation}
where the coefficients are normalized Bessel coefficients,
\begin{equation}
c_0 = \frac{J_0(\beta)}{\sqrt{J^2_0(\beta)+J^2_1(\beta)}} \hspace{0.3cm}\text{and}\hspace{0.3cm} c_1 = \frac{J_1(\beta)}{\sqrt{J^2_0(\beta)+J^2_1(\beta)}}e^{i\pi/2}
\label{eq:outfield5}
\end{equation}
The Bessel coefficients are a function of the applied microwave voltage. By changing the voltage and the phase of the microwave field, one can rotate the qubit to an arbitrary point on the Bloch sphere. Similarly, the first two sidebands and the carrier can be used as a qutrit for implementing frequency coded BB84 protocols.

%